\documentclass[aps,prl,reprint,superscriptaddress,nofootinbib]{revtex4-2}

\usepackage{amsmath,amssymb,bm,mathtools,xcolor}
\usepackage{graphicx}
\usepackage{xspace}
\usepackage{hyperref}
\hypersetup{colorlinks=true,linkcolor=blue,citecolor=blue,urlcolor=blue}
\usepackage{tikz}
\usetikzlibrary{shapes.geometric,decorations.pathmorphing,decorations.markings,arrows.meta,calc}

\newcommand{\cH}{\mathcal{H}}
\newcommand{\cK}{\mathcal{K}}
\newcommand{\cN}{\mathcal{N}}

\newcommand{\cO}{\mathcal{O}}
\newcommand{\cV}{\mathcal{V}}

\newcommand{\ord}{\mathcal{O}}
\newcommand{\dd}{\mathrm{d}}
\newcommand{\ee}{\mathrm{e}}
\newcommand{\ii}{\mathrm{i}}
\newcommand{\eps}{\varepsilon}
\newcommand{\wt}[1]{\widetilde{#1}}
\newcommand{\DDVCS}{\textsc{DDVCS}\xspace}
\newcommand{\DVCS}{\textsc{DVCS}\xspace}

\newcommand{\hH}{\widehat{\cH}}

\begin{document}

\title{From Vacuum to Nucleon: Fixed--$j$ Kernel Matching of Holographic Current Correlators to QCD}

\author{Kiminad A.~Mamo}
\affiliation{Department of Physics, University of Connecticut, Storrs, CT 06269-3046, USA}
\email{ska25005@uconn.edu}
\date{\today}

\begin{abstract}
We show that the normalized non-normalizable photon profile fixed by the vacuum vector-current two-point function, when inserted into the fixed-$j$ hadronic Witten diagram, produces the fixed-$j$ conformal hard kernel in the leading-twist singlet vector channel of deeply virtual Compton scattering (DVCS) and double deeply virtual Compton scattering (DDVCS). For the near-boundary Brower--Polchinski--Strassler--Tan closed-string upper vertex, the Witten diagram yields the Gauss-hypergeometric Wilson-kernel family of the QCD conformal OPE, with the Mellin exponent derived from the explicit $z$-power count \(\delta_c(j)=j+\Delta_c(j)-2\), not inserted as an ansatz. The lower Witten vertex is matched to, or provides a holographic representation of, the nonperturbative conformal moment; it is not a new hard kernel or a freely adjustable skewness profile. The matching is TT-projected, kernel-level, and fixed-scale in the conformal partial-wave/CS representation: finite coefficient normalizations, spin-$j$ source normalizations, scale factors, and lower hadronic matrix elements remain matched inputs. Together with the Nishio--Watari open branch and the protected/unprotected \(j=2\) split, this structurally identifies the protected closed branch with the $(-)$ conformal partial wave and the even open branch with the unprotected $(+)$ counterpart in the projected fixed-$j$ singlet vector amplitude; these are mixed singlet eigenchannels, not literal unmixed quark/gluon operators at finite $N_c$. The conformal/Gegenbauer basis used in generalized parton distribution (GPD) deconvolution therefore has a second ultraviolet origin: it is selected both by the QCD OPE and by the near-boundary AdS Witten vertex.
\end{abstract}

\maketitle

The classic ultraviolet test of bottom-up holographic QCD is the vacuum vector-current two-point function. In AdS/QCD, the on-shell action of a bulk gauge field reproduces the logarithmic $Q^2$ behavior of the QCD current-current correlator and fixes the bulk gauge coupling~\cite{Erlich:2005qh}; in the soft-wall model the same current is sourced by the Grigoryan--Radyushkin bulk-to-boundary propagator~\cite{Grigoryan:2007vg}. The vacuum computation fixes the normalized non-normalizable current source. It does not by itself fix the spin-$j$ exchange normalization, trajectory data, or lower hadronic matrix element; those remain matched or modeled inputs. This Letter asks what happens when that same source is inserted into the fixed-$j$ hadronic Witten diagram. The answer is: the vacuum-normalized photon profile produces the same fixed-$j$ conformal Wilson-kernel family inside a hadron, while the lower Witten vertex is matched to the nonperturbative conformal moment. The novelty is not the open \(C_1\) identity itself, see Eq.~\ref{eq:C1_integral}. It is that the closed BPST fixed-\(j\) upper Witten vertex independently produces the same conformal Wilson-kernel family, fixes the Mellin label by \(z\)-power counting, and thereby completes the structural open/closed \(\leftrightarrow(+)/(-)\) operator-basis mapping relevant for \DDVCS/\DVCS deconvolution. In this organization, the hadronic Compton amplitude is written as conformal moments multiplying the same hard kernel selected by the QCD OPE, rather than as a freely chosen $x$-space skewness profile.

Deeply virtual Compton scattering (DVCS) probes a short-distance current-current correlator inside a strongly bound hadron, and double deeply virtual Compton scattering (DDVCS) keeps both photons off shell. At leading power, \DVCS and \DDVCS separate this hard current kernel from long-distance hadronic matrix elements. This is the content of the collinear factorization theorem in the generalized Bjorken regime~\cite{Ji:1998xh,Collins:1998be}. The same amplitude may be written in the original momentum-fraction language of GPDs, introduced by M\"uller \textit{et al.}, by Ji and by Radyushkin~\cite{Muller:1994ses,Ji:1996ek2,Ji:1996ek,Radyushkin:1996ru,Radyushkin:1996nd}, or in the conformal-basis language, where complex conformal spin organizes local twist-two operators and diagonalizes the leading-order singlet evolution~\cite{Mueller:1997hs,Belitsky:1997rh,Mueller:2005ed,Kumericki:2006xx,Kumericki:2007sa}. This matters for JLab, CLAS/SoLID, and future EIC extractions because the quantities to constrain are physical conformal moments multiplying a fixed kernel, not arbitrary $x$-space skewness profiles. Recent two-loop work has made this conformal-basis description increasingly explicit, including vector, axial-vector, transversity, \DDVCS, and conformal-moment coefficient functions~\cite{Braun:2020yib,Braun:2021ysj,Braun:2022nnlo,Ji:2023transversity,Braun:2024ddvcs,Braun:2026dvcsmom}. Recent \DDVCS phenomenology and sensitivity studies emphasize the same kinematic motivation: the second photon virtuality opens access to GPD information beyond the on-shell \DVCS/timelike Compton scattering (TCS) lines, with feasibility studied for Jefferson Lab/CEBAF, CLAS/SoLID, and EIC settings~\cite{Deja:2023ahc,Alvarado:2025prc,Alvarado:2026ggy,Accardi:2023chb}. String/holographic trajectories are useful here not as new fit functions for skewness, but because they give a fixed-$j$ Witten-diagram realization of the same operator moments that enter this conformal basis.

This ultraviolet-to-hadron extension also addresses the deconvolution problem directly. In momentum-fraction space, extracting GPDs from \DVCS/\DDVCS observables is underconstrained: many $x$-space profiles can project onto nearly the same Compton form factor~\cite{Bertone:2021yyz}. Conformal moments reorganize the same information in terms of local twist-two operators selected by Lorentz symmetry, so their skewness dependence obeys polynomiality by construction. Holographic QCD adds a second reason for using this basis. The lower Witten vertex models the same fixed-$j$ conformal moments geometrically---model dependently, but with polynomiality constraints kept explicit for the physical even-spin local moments. The same framework has been employed in the parametrization and global extraction of small-skewness GPDs~\cite{Guo:2022ump0,Guo:2023ump,Guo:2025muf}, and in all-skewness string/holographic conformal-moment GPD parametrizations~\cite{Mamo:2024jwp,Mamo:2024vjh,Hechenberger:2025rye,Hechenberger:2025wnz}. Once a holographic-string construction provides such constrained conformal-moment data, the inverse problem is not the unconstrained reconstruction of arbitrary $x$-space profiles; it is the determination of physical conformal moments multiplying a fixed Wilson kernel.

The central claim is therefore sharper than the statement that holographic QCD can model a Compton amplitude. Before any infrared model of the nucleon is used, the ultraviolet Witten vertex already produces the same fixed-$j$ conformal Wilson-kernel family as the QCD hard-current OPE. The lower vertex is then matched to the nonperturbative conformal moment. No phenomenological fit or infrared ansatz is used to obtain the hard kernel; the infrared model enters only through the lower conformal moment. This is the hadronic-state analogue of the vacuum current matching in the following fixed-$j$ sense: the vacuum computation fixes the current source, while the hadronic Witten diagram shows that this same source yields the QCD conformal Wilson-kernel family. The exchanged-operator normalization and lower conformal moment remain matched inputs.

At high energy and small $x$, the closed-string channel is anchored by the Brower--Polchinski--Strassler--Tan (BPST) graviton trajectory~\cite{Brower:2006ea}, while holographic \DVCS amplitudes have been analyzed in conformal Regge theory by Costa and Djuri\'c, in related work by Brower \textit{et al.}~\cite{Costa:2012cb,Brower:2012mk}, and on $s$-channel/large-$x$ scalar target in~\cite{Marquet:2010sf}. For the open Reggeon construction, Nishio--Watari already displayed the $C_1(\delta,\vartheta)$ AdS integral and its Gauss-hypergeometric form in Appendix~B, especially Eqs.~(285)--(289) of~\cite{Nishio:2014rya,Nishio:2014eua}. The novelty contrast is thus sharply delimited. Nishio--Watari exposed the open-string $C_1$ hypergeometric kernel. The content here is that the same family appears in the closed BPST channel, with the Mellin exponent fixed by an independent near-boundary $z$-power count that uses the BPST boundary mode and trajectory data and has no open-string input. The lower vertex is then matched to the nonperturbative conformal moment. The protected $j=2$ split identifies the protected closed branch and supplies the analytic-continuation prescription used for the structural closed/open $\leftrightarrow$ $(-)/(+)$ fixed-$j$ dictionary.

The hadronic object is the off-forward matrix element of two electromagnetic currents,
\begin{equation}
T^{\mu\nu}(P,\Delta,q)=
\ii\!\int\!\dd^4x\,\ee^{-\ii q\cdot x}
\langle P'|\,T\{J^\mu(x/2)J^\nu(-x/2)\}\,|P\rangle,
\label{eq:compton-tensor}
\end{equation}
which is probed in \DDVCS and, after taking one photon on shell, in \DVCS; a corresponding Witten diagram is sketched in Fig.~\ref{fig:vacuum_to_hadron}. The statement concerns the leading-twist unpolarized singlet vector invariant amplitude in the collinear window below. It is not a claim about the full tensor decomposition including all helicity, axial, transversity, contact, target-mass, or higher-twist structures. This Letter isolates the fixed-$j$ kernel statement; the accompanying long paper~\cite{Mamo:2026fjh} gives the full derivation, normalization conventions, open-channel construction, running-coupling bookkeeping, and phenomenological implementation details.

\begin{figure*}[t]
\centering
\tikzset{
  boldphoton/.style={decorate, decoration={snake, segment length=2mm, amplitude=0.5mm}, draw=black, line width=0.4mm},
  particle/.style={draw=black, postaction={decorate}, decoration={markings, mark=at position .55 with {\arrow{stealth}}}},
  antiparticle/.style={draw=black, postaction={decorate}, decoration={markings, mark=at position .55 with {\arrowreversed{stealth}}}},
  dbl_wiggly/.style={double, decorate, decoration={snake, amplitude=1pt, segment length=3mm}, double distance=0.55pt, line width=0.8pt},
  hexagram/.style={draw, star, star points=6, star point ratio=0.5, fill=black, minimum size=1.2mm, inner sep=0pt}
}

\begin{minipage}[t]{0.34\textwidth}
\centering
\begin{tikzpicture}[scale=0.78, transform shape]
  \node (J1) at (0,4) [hexagram, label={[align=center]above left:$J^\mu(x;0)$}] {};
  \node (J2) at (4,4) [hexagram, label={[align=center]above right:$J^\nu(0;0)$}] {};
  \draw[thick] (2,4) circle (2);
  \node[red,font=\normalsize] at ($(J1)+(-0.45,-0.18)$) {$\frac{1}{g_5}$};
  \node[red,font=\normalsize] at ($(J2)+(0.45,-0.18)$) {$\frac{1}{g_5}$};
  \draw[boldphoton] (J1) -- (J2)
      node[midway, above, sloped] {$z^{-1}\partial_z\cV(q,z=\eps)$};
\end{tikzpicture}

\vspace{0.35ex}
{\small (a) Vacuum current-current correlator}
\end{minipage}
\hfill
\begin{minipage}[t]{0.62\textwidth}
\centering
\begin{tikzpicture}[scale=0.66, transform shape]
    \pgfmathsetmacro{\Rc}{sqrt(8)}

    \coordinate (P1c) at (0,0);
    \coordinate (J1c) at (0,4);
    \coordinate (P2c) at (4,0);
    \coordinate (J2c) at (4,4);

    \draw[thick] (2,2) circle[radius=\Rc cm];

    \node[hexagram, label={[font=\scriptsize,align=center]below left:$\cO_{P}(y_1;0)$}] (P1) at (P1c) {};
    \node[hexagram, label={[font=\scriptsize,align=center]above left:$J^\mu(x_1;0)$}] (J1) at (J1c) {};
    \node[hexagram, label={[font=\scriptsize,align=center]below right:$\bar{\cO}_{P}(y_2;0)$}] (P2) at (P2c) {};
    \node[hexagram, label={[font=\scriptsize,align=center]above right:$J^\nu(x_2;0)$}] (J2) at (J2c) {};

    \node[fill, circle, inner sep=2pt] (vb) at (2,1) {};
    \node[fill, circle, inner sep=2pt] (vt) at (2,3) {};

    \draw[particle]     (P1) -- (vb) node[pos=0.54, below, sloped, font=\scriptsize] {$\Psi_{P}(p_1;z')$};
    \draw[antiparticle] (P2) -- (vb) node[pos=0.46, below, sloped, font=\scriptsize] {$\bar\Psi_{P}(p_2;z')$};
    \draw[boldphoton]   (J1) -- (vt) node[pos=0.52, above, sloped, font=\scriptsize] {$\cV(q_1;z)$};
    \draw[boldphoton]   (vt) -- (J2) node[pos=0.48, above, sloped, font=\scriptsize] {$\cV(q_2;z)$};
    \draw[dbl_wiggly]   (vt) -- (vb) node[midway, right, font=\scriptsize] {$G_j^{(X)}(\Delta;z,z')$};

    \node[font=\scriptsize] at (2.23,2.92) {$z$};
    \node[font=\scriptsize] at (2.26,1.10) {$z'$};
    \node[font=\scriptsize] at (2,4.30) {$T\{J^\mu J^\nu\}$};
    \node[font=\scriptsize] at (2,-0.3) {$\langle P'|\cdots|P\rangle$};

    \node[red,font=\normalsize] at ($(J1)+(-0.38,-0.08)$) {$\frac{1}{g_5}$};
    \node[red,font=\normalsize] at ($(J2)+(0.38,-0.08)$) {$\frac{1}{g_5}$};
\end{tikzpicture}

\vspace{0.35ex}
{\small (b) Current-current correlator inside a hadron}
\end{minipage}
\caption{(a) Vacuum current-current correlator, whose ultraviolet logarithm fixes $g_5$ by matching to the leading QCD vector-current two-point function~\cite{Erlich:2005qh}. (b) Unfactorized off-forward Witten diagram for the hadronic current-current correlator relevant for \DDVCS/\DVCS. The two currents couple through bulk photon fields to a spin-$j$ exchange $G_j^{(X)}$, which in turn couples to the incoming and outgoing nucleon wave functions. The $1/g_5$ labels mark the same non-normalizable current insertions as in panel (a); other normalization factors are suppressed. Fixed-$j$ factorization operationally isolates the upper photon blob evaluated in Eqs.~\eqref{eq:C1_integral}--\eqref{eq:C1_kernel_form} and the lower hadronic conformal moment $\Phi_N^{(X)}$ in Eq.~\eqref{eq:holo_fixedj}.}
\label{fig:vacuum_to_hadron}
\end{figure*}

The relevant collinear window is
\begin{align}
&Q_1^2\sim Q_2^2\sim \wt Q^{\,2}\equiv Q^2,
\qquad
-t,M_N^2\ll Q_i^2,\nonumber\\
&\xi,\eta\ll 1,
\qquad
\vartheta\equiv\frac{\eta}{\xi}=\mathcal O(1),
\qquad i=1,2.
\label{eq:window}
\end{align}
Within this window the matching is performed after applying the transverse--transverse (TT) scalar projector to the current correlator. Gauge-invariant completions of the $VV\cO_j$ upper vertex can generate $F_{z\mu}$, $K_0$, or $z\partial_zK_1$ structures, but these project onto longitudinal or distinct invariant amplitudes and are not part of the scalar kernel matched here. Equivalently, for the leading scalar upper vertex,
\begin{equation}
-\frac12 g^T_{\mu\nu}\,\cV^{\mu\nu}_{VV\cO_j}
=\cN_j\,\cV(Q_1,z)\cV(Q_2,z)
+\ord\!\left(\frac{M_N^2}{Q^2},\frac{-t}{Q^2}\right),
\label{eq:prl-tt-projection}
\end{equation}
while the $F_{z\mu}$-induced pieces feed LL, TL, or helicity-distinct projections. No statement is made here about those LL/TL or helicity-distinct kernels; they require the derivative/$F_{z\mu}$ structures projected out of the scalar TT invariant. This is the same TT/LL separation used in the Nishio--Watari construction and in the holographic electroproduction analysis of Sec.~II.D, especially Eq.~(II.17), of Ref.~\cite{Mamo:2021tzd}.
Here $\sim$ denotes comparable hard virtualities, not equality of the two photon virtualities. The off-forward variables are
\begin{equation}
\eta=\frac{\Delta\!\cdot\!\wt q}{2P\!\cdot\!\wt q},
\qquad
\xi=\frac{\wt Q^2}{2P\!\cdot\!\wt q},
\qquad
\wt q=\frac{q_1+q_2}{2}.
\end{equation}
\DDVCS is the natural arena because the ratio $\eta^2/\xi^2$ remains fully active only when both photons are off shell. The \DVCS limit is obtained only after the fixed-$j$ kernel has been identified.

It is useful to name the universal conformal kernel once:
\begin{equation}
\cK_j^{(\gamma)}(\vartheta)
\equiv
{}_2F_1\!\left(
\frac j2+\frac\gamma4,
\frac{j+1}{2}+\frac\gamma4;
j+\frac32+\frac\gamma2;
\vartheta^2
\right).
\label{eq:universal_kernel}
\end{equation}
Throughout, ``Wilson kernel'' denotes this hypergeometric functional dependence; ``Wilson coefficient'' denotes the full normalized coefficient, including finite factors such as $c_j^\pm$ and scale/evolution factors.

In the accompanying long paper~\cite{Mamo:2026fjh}, the fixed-$j$ holographic amplitudes are derived as
\begin{align}
\hH_{\rm holo}^{(X)}(j)
&=\xi^{-j}\left(\frac{\mu}{Q}\right)^{\gamma_X(j)}
\cK_j^{(\gamma_X(j))}(\vartheta)\nonumber\\
&\quad\times
\left(\frac{\mu_0}{\mu}\right)^{\gamma_X(j)}
\Phi_N^{(X)}(j;t,\eta),
\label{eq:holo_fixedj}
\end{align}
with $X=o,c$ denoting the even open and closed channels realized, respectively, by the Nishio--Watari open-string and BPST closed-string trajectories. Here $\Phi_N^{(X)}$ denotes the channel-dependent nonperturbative conformal-moment factor, including the lower Witten vertex and the fixed normalization conventions used to compare with the QCD conformal moment. The key factorization point is that the upper Witten vertex carries the entire hard dependence on $Q^2$ and $\eta/\xi$, while $\Phi_N^{(X)}$ carries the hadronic matrix element. This is the precise intersection with the GPD deconvolution problem: the holographic input is a conformal moment multiplying a fixed Wilson kernel, not a freely adjustable $x$-space skewness profile.

The universal kernel is fixed by the upper bulk integral
\begin{align}
C_1(\delta,\vartheta)
&=(1-\vartheta^2)^{1/2}
\int_0^\infty\!\dd y\,y^{1+\delta}\nonumber\\[-0.3ex]
&\quad\times
K_1\!\left(y\sqrt{1+\vartheta}\right)
K_1\!\left(y\sqrt{1-\vartheta}\right),
\label{eq:C1_integral}
\end{align}
whose closed form is
\begin{equation}
C_1(\delta,\vartheta)=N(\delta)
{}_2F_1\!\left(
\frac{\delta}{4},\frac{\delta}{4}+\frac12;
\frac{\delta}{2}+\frac32;\vartheta^2\right).
\label{eq:C1_kernel_form}
\end{equation}
The normalization $N(\delta)$ is displayed in the long paper; here only the functional dependence is needed. The displayed square-root representation is the spacelike \DDVCS form with $|\vartheta|<1$; physical endpoint limits and complex continuations are taken only after the fixed-$j$ kernel has been identified, with the usual hard-scattering prescription. The Mellin exponent is not assumed but fixed by the Witten-diagram power counting,
\begin{equation}
\delta_X(j)=j+\Delta_X(j)-2=2j+\gamma_X(j).
\label{eq:delta}
\end{equation}
Here $\Delta_X(j)=2+j+\gamma_X(j)$ is the dimension of the exchanged spin-$j$ mode in channel $X$.
Explicitly, the upper-vertex powers give
\begin{equation}
 z^{-5}\,z^{4+2(j-2)}\,z^{-(j-2)}\,z^{\Delta_X(j)}\,z^2
 =z^{j+\Delta_X(j)-1}=z^{1+\delta_X(j)}.
\label{eq:prl-z-count}
\end{equation}
The factors in Eq.~\eqref{eq:prl-z-count} are, respectively, the AdS measure, the spin-$j$ tensor vertex, the propagator factor, the boundary mode, and the two photon wave functions. Equations~\eqref{eq:C1_integral}--\eqref{eq:prl-z-count} are the hadronic-state analogue of the vacuum matching in a restricted sense: the normalized current insertion that reproduces the vacuum logarithm now produces the universal conformal Wilson-kernel family of the off-forward current-current correlator inside a hadron. The hypergeometric form is compatible with conformal kinematics; the nontrivial holographic input is that the fixed-$j$ Witten diagram fixes the Mellin label $\delta_X(j)=j+\Delta_X(j)-2$ from the actual near-boundary $z$ powers and that the $j=2$ protected branch fixes the structural channel assignment. For $X=c$, this is the closed BPST derivation emphasized here: the exponent $\delta_c(j)$ comes from the actual $z$ powers in the upper Witten vertex. For $X=o$, Appendix~B of Nishio--Watari gives the same $C_1$ integral and hypergeometric transformation in Eqs.~(285)--(289) for the open Reggeon construction~\cite{Nishio:2014rya,Nishio:2014eua}. Crucially, the open branch is not inferred from the closed formula. The open Nishio--Watari upper vertex has its own fixed-$j$ Witten representation; only after that parallel factorized vertex is written is the replacement notation used as shorthand. Within QCD, collinear factorization is a theorem for the same hard-current object. Within the fixed-$j$ semiclassical holographic expansion, the ultraviolet impact factor is directly derived from the Witten diagram and matches the QCD conformal-OPE hard kernel. The matching below is therefore a kernel-level fixed-scale statement, followed by matching of the lower conformal moments.

In the singlet conformal basis that diagonalizes the LO evolution, and using the physical-spin convention related to Ref.~\cite{Kumericki:2007sa} by $j_{\rm ref}=j-1$, the fixed-$j$ QCD amplitude is displayed in the conformal-OPE/conformal-subtraction representation as
\begin{align}
\hH_{\rm pQCD}^{\rm sing}(j)
&=\sum_{a=\pm}
 \xi^{-j}\,c_j^a(\alpha_s^\ast)
\left(\frac{\mu}{Q}\right)^{\gamma_j^a(\alpha_s^\ast)}\nonumber\\[-0.3ex]
&\quad\times
\cK_j^{(\gamma_j^a(\alpha_s^\ast))}(\vartheta)
\left(\frac{\mu_0}{\mu}\right)^{\gamma_j^a(\alpha_s^\ast)}
H_j^a(\eta,t;\mu_0).
\label{eq:pqcd_fixedj}
\end{align}
In Eq.~\eqref{eq:pqcd_fixedj}, the displayed powers are the conformal-limit representation of the Wilson-coefficient and conformal-moment evolution factors. At the matching point below one may read $\alpha_s^\ast$ as the coupling evaluated at $\mu_\ast$ in this representation. The finite normalizations $c_j^a$ are not set to one; they remain part of the matched fixed-scale coefficient.

The comparison with Eq.~\eqref{eq:holo_fixedj} is immediate. At the single matching scale
\begin{equation}
Q=\mu=\mu_0=\mu_\ast,
\label{eq:mu_star}
\end{equation}
the channel-by-channel dictionary for the anomalous-dimension parameter appearing in the fixed-$j$ kernel is
\begin{equation}
\gamma_o(j)\ \stackrel{\rm kernel}{\longleftrightarrow}\ 
\gamma_j^+(\alpha_s^\ast),
\qquad
\gamma_c(j)\ \stackrel{\rm kernel}{\longleftrightarrow}\ 
\gamma_j^-(\alpha_s^\ast),
\label{eq:dictionary}
\end{equation}
together with
\begin{align}
\Phi_N^{(o)}(j;t,\eta)&\leftrightarrow c_j^+H_j^+(\eta,t;\mu_\ast),\nonumber\\
\Phi_N^{(c)}(j;t,\eta)&\leftrightarrow c_j^-H_j^-(\eta,t;\mu_\ast).
\label{eq:moments}
\end{align}
Equations~\eqref{eq:dictionary} and \eqref{eq:moments} are the fixed-$j$ structural matching statement of the Letter. The arrows in Eq.~\eqref{eq:dictionary} are a dictionary for the conformal-dimension label inside the fixed-$j$ kernel, not a numerical equality between weak-coupling QCD anomalous dimensions and strong-coupling holographic trajectory functions as functions of $j$. The equality $Q=\mu=\mu_0=\mu_\ast$ is a matching choice for renormalized fixed-$j$ coefficients and conformal moments; it is not a claim that the holographic infrared wave functions live at the hard scale. The nontrivial point is not that equal-endpoint evolution operators become identities. It is that, before the lower moment is matched, the ultraviolet Witten vertex has already produced the same Gauss-hypergeometric Wilson-kernel family as a function of $\eta/\xi$, $j$, and the anomalous-dimension parameter. At the matching scale, the finite forward normalizations $c_j^\pm$ and the nonperturbative conformal moments remain; the lower factors in Eq.~\eqref{eq:moments} are matched rather than predicted from the UV kernel.

Equivalently, the matching is a fixed-scale identification of the fixed-$j$ hard kernel, supplemented by the fixed-$j$ open/closed $\leftrightarrow$ $(+)/(-)$ channel assignment. In the conformal partial-wave/CS representation of the projected invariant used here, the matched object is the fixed-$j$ conformal kernel rather than an all-scale equality between full theories. Beta-proportional conformal-anomaly terms, scheme transformations, and logarithmic running are part of the coefficient/evolution bookkeeping away from the single matching point and lie outside the kernel-level statement. Higher-order perturbative information enters the matched fixed-$j$ coefficient through $\gamma_j^\pm(\alpha_s)$ and the finite normalizations $c_j^\pm(\alpha_s(\mu_\ast))$. At the matching point, this information selects the anomalous-dimension label and finite forward normalization of the same conformal partial wave, not an alternative fixed-scale $\eta/\xi$ Wilson-kernel family. For fits at intermediate experimental $Q^2$, the scale $\mu_\ast$ should be treated as the input matching scale for the conformal moments. The matched moments are then evolved to the measured scale, while target-mass, higher-twist, and genuinely noncollinear corrections are added to the projected leading-twist description rather than absorbed into a new fixed-$j$ $\eta/\xi$ hard kernel. The accompanying long paper gives the explicit running-coupling formulas, normalization details, and explicit open-channel construction and BPST--Nishio--Watari open/closed map~\cite{Mamo:2026fjh}.

The channel assignment is anchored dynamically by the first physical even moment. In QCD, the anomalous dimensions
\begin{equation}
\gamma_{2}^{-,(0)}=0,
\qquad
\gamma_{2}^{+,(0)}=\frac{4}{3}\left(2C_F+T_F n_f\right),
\label{eq:j2_qcd}
\end{equation}
so the $(-)$ eigenchannel is protected by the singlet momentum-sum rule, whereas the $(+)$ channel is not. In holography,
\begin{equation}
\Delta_c(2)=4,
\qquad
\gamma_c(2)=0,
\qquad
\gamma_o(2)=\sqrt{1+\sqrt\lambda}-2,
\label{eq:j2_holo}
\end{equation}
so the closed branch passes through the bulk graviton and is protected by the conserved total energy--momentum tensor, independent of the value of $\lambda$ within the BPST trajectory. The open and closed labels are therefore effective diagonal branches in the projected large-$N_c$/holographic description, not literal unmixed quark and gluon operators at finite $N_c$; the QCD $\pm$ moments are quark--gluon mixtures. The even open value is $\lambda$ dependent and unprotected; in the large-$\lambda$ regime it is finite and positive. At the accidental zero, such as the formal $\lambda=9$ zero of Eq.~\eqref{eq:j2_holo}, the open and closed values are temporarily degenerate at $j=2$, but the open branch remains unprotected for $j\neq2$ and the full trajectory structure, not the isolated value at one coupling, carries the unprotected assignment. The same distinction appears in the finite-$\lambda$ trajectory data,
\begin{align}
\gamma_c(j)&=-j+\sqrt{4+2\sqrt\lambda\,(j-2)},\\
\gamma_o(j)&=-j+\sqrt{1+\sqrt\lambda\,(j-1)},
\label{eq:branches}
\end{align}
with branch points $j_{0c}=2-2/\sqrt\lambda$ and $j_{0o}=1-1/\sqrt\lambda$. Thus the common shorthand $\gamma_c\sim\sqrt{j-2}$ and $\gamma_o\sim\sqrt{j-1}$ should be read as the strong-coupling/intercept diagnostic of the protected closed branch and the unprotected open branch, not as a statement that the finite-$\lambda$ branch points sit exactly at $2$ and $1$. At fixed $j$, both for physical even spins and for the complex-$j$ continuation used in Regge reconstruction, the protected closed branch is structurally identified with the $(-)$ conformal partial wave, while the even open branch gives the unprotected $(+)$ counterpart in the projected fixed-$j$ singlet vector amplitude. Within the Nishio--Watari open-sector realization, the open Reggeon construction supplies the open-sector fixed-$j$ kernel; the singlet $(+)$ assignment follows from the $j=2$ protected/unprotected structure and the fixed-scale matching dictionary. The replacement rule used in the companion derivation is only a notation-saving map between parallel fixed-$j$ factorized expressions, not a derivation of the full open-string sector from the closed one.

The message of the Letter can now be stated in one line. The novelty is not the open \(C_1\) identity itself; it is that the closed BPST fixed-\(j\) upper Witten vertex independently produces the same conformal Wilson-kernel family, fixes the Mellin label by \(z\)-power counting, and thereby completes the structural open/closed \(\leftrightarrow(+)/(-)\) operator-basis mapping relevant for \DDVCS/\DVCS deconvolution. The vacuum two-point function fixes the normalization of the holographic current source by reproducing the leading QCD logarithm~\cite{Erlich:2005qh}; the spin-$j$ exchange normalization, trajectory data, and lower hadronic matrix element remain matched inputs. The hadronic current-current correlator inherits that source and, after fixed-$j$ factorization, matches the leading-twist singlet vector conformal Wilson-kernel family of QCD at one scale. The IR conformal moments are matched nonperturbative inputs, subject to the usual constraints such as polynomiality and Lorentz covariance, rather than quantities predicted by the UV matching itself.

This gives a concrete physics reason why the conformal/Gegenbauer basis is privileged for the \DVCS/\DDVCS deconvolution problem~\cite{Bertone:2021yyz}. More than a convenient orthogonal expansion, it is the basis in which the QCD hard-current OPE, leading-order singlet evolution, skewness polynomiality, and the near-boundary AdS Witten vertex all expose the same fixed-$j$ structure. This matters directly for JLab/CLAS/SoLID and future EIC analyses: the quantity to constrain is a set of physical conformal moments multiplying a fixed hard kernel, not an arbitrary $x$-space skewness profile. Holographic QCD thus reproduces not only the familiar vacuum current correlator, but also the universal fixed-$j$ hard kernel of a current-current correlator inside a hadron. This is the precise sense in which holographic \DDVCS/\DVCS becomes a hadronic generalization of the classic vacuum matching: the upper vertex fixes the conformal Wilson-kernel family, and the lower vertex provides the matched holographic representation of the conformal moment.

\begin{acknowledgments}
I thank Christian Weiss for discussions. I also thank Jefferson Lab for hospitality during the completion of this work. This work was supported by DOE grant no. DE-FG02-04ER41309, NSF grant no. 2412625, and DOE under the umbrella of the Quark-Gluon Tomography (QGT) Topical Collaboration with Award No. DE-SC0023646.
\end{acknowledgments}

\bibliography{ddvcs_prl_refs}

\end{document}